\documentclass[12pt,nofootinbib]{article}

\usepackage{graphicx}
\usepackage{amssymb}
\usepackage{amsmath}
\usepackage{color}
\usepackage[colorlinks=true,linkcolor=blue,citecolor=blue]{hyperref}

\begin{document}
\renewcommand{\thefootnote}{\#\arabic{footnote}}
\newcommand{\rem}[1]{{\bf [#1]}} \newcommand{\gsim}{ \mathop{}_
{\textstyle \sim}^{\textstyle >} } \newcommand{\lsim}{ \mathop{}_
{\textstyle \sim}^{\textstyle <} } \newcommand{\vev}[1]{ \left\langle
{#1} \right\rangle } \newcommand{\bear}{\begin{array}} \newcommand
{\eear}{\end{array}} \newcommand{\bea}{\begin{eqnarray}}
\newcommand{\eea}{\end{eqnarray}} \newcommand{\beq}{\begin{equation}}
\newcommand{\eeq}{\end{equation}} \newcommand{\bef}{\begin{figure}}
\newcommand {\eef}{\end{figure}} \newcommand{\bec}{\begin{center}}
\newcommand {\eec}{\end{center}} \newcommand{\non}{\nonumber}
\newcommand {\eqn}[1]{\beq {#1}\eeq} \newcommand{\la}{\left\langle}
\newcommand{\ra}{\right\rangle} \newcommand{\ds}{\displaystyle}
\newcommand{\red}{\textcolor{red}} 
\def\SEC#1{Sec.~\ref{#1}} \def\FIG#1{Fig.~\ref{#1}}
\def\EQ#1{Eq.~(\ref{#1})} \def\EQS#1{Eqs.~(\ref{#1})} \def\lrf#1#2{
\left(\frac{#1}{#2}\right)} \def\lrfp#1#2#3{ \left(\frac{#1}{#2}
\right)^{#3}} \def\GEV#1{10^{#1}{\rm\,GeV}}
\def\MEV#1{10^{#1}{\rm\,MeV}} \def\KEV#1{10^{#1}{\rm\,keV}}
\def\REF#1{(\ref{#1})} \def\lrf#1#2{ \left(\frac{#1}{#2}\right)}
\def\lrfp#1#2#3{ \left(\frac{#1}{#2} \right)^{#3}} \def\OG#1{ {\cal
O}(#1){\rm\,GeV}}

\begin{titlepage}

\begin{flushright}
ICRR-Report-692-2014-18 \\
IPMU-14-0301 \\
RUP-14-14 
\end{flushright}

\begin{center}

\vskip 1.2cm

{\usefont{T1}{pag}{m}{n}
{\Large CDM/baryon isocurvature perturbations \\[0.5em]
in a sneutrino curvaton model} 
}

\vskip 1.2cm

{\usefont{T1}{bch}{m}{n}
Keisuke Harigaya$^a$, Taku Hayakawa$^b$, \\[0.3em]
Masahiro Kawasaki$^{a,b}$
and
Shuichiro Yokoyama$^{c,b}$}

\vskip 0.4cm

{\it$^a$Kavli IPMU (WPI), TODIAS, The University of Tokyo, 5-1-5 Kashiwanoha, 
Kashiwa, 277-8583, Japan}\\
{\it$^b$Institute for Cosmic Ray Research, The University of Tokyo,
5-1-5 Kashiwanoha, Kashiwa, Chiba 277-8582, Japan}\\
{\it$^c$Department of Physics, Rikkyo University,
3-34-1 Nishi-Ikebukuro, Toshima, Tokyo 223-8521, Japan}

\date{\today}

\vskip 1.2cm

\begin{abstract}
Matter isocurvature perturbations are strictly constrained 
from cosmic microwave background observations. 
We study a sneutrino curvaton model 
where both cold dark matter (CDM)/baryon isocurvature perturbations are generated. 
In our model, total matter isocurvature perturbations are reduced
since the CDM/baryon isocurvature perturbations compensate for each other. 
We show that this model can not only avoid the stringent observational constraints 
but also suppress temperature anisotropies on large scales, 
which leads to improved agreement with observations.

\end{abstract}

\end{center}
\end{titlepage}

\baselineskip 6mm

\section{Introduction}
\label{sec:intro}

Inflation~\cite{Guth:1980zm,Sato:1980yn,Starobinsky:1980te,Kazanas:1980tx,%
Linde:1981mu,Albrecht:1982wi,Linde:1983gd} 
in the very early universe solves horizon and flatness problems in the standard 
big bang cosmology.
Furthermore, light scalar fields acquire fluctuations during inflation 
and can generate adiabatic and almost scale-invariant density perturbations 
which are in good agreement with cosmic microwave background (CMB) 
observations~\cite{Smoot:1992td,Hinshaw:2012aka,Ade:2013uln}.
In the simplest case, a scalar field which causes inflation (called an inflaton)
is responsible for the density perturbations,
but it is also possible that another scalar field gives a significant
contribution to the density perturbations. 
In particular, in curvaton models~\cite{Enqvist:2001zp,Lyth:2001nq,Moroi:2001ct} 
some scalar field besides the inflaton (called a curvaton) obtains fluctuations during inflation 
and decays into radiation to produce adiabatic perturbations 
after reheating due to the inflaton decay. 

However, in general, curvaton models produce not only adiabatic perturbations 
but also isocurvature ones. 
Isocurvature perturbations $S_{ij}$ are defined as
\begin{equation}
   S_{ij} = \frac{\delta\rho_i}{(1+w_i)\rho_i}-\frac{\delta\rho_j}{(1+w_j)\rho_j},
\end{equation}
where $\rho_i$, $\delta\rho_i$ and $w_i$ are the energy density, its fluctuation and 
the coefficient of the equation of state of a component $i$. 
For example, if cold dark matter (CDM) (or baryon number) is generated and decouples from 
thermal bath before the curvaton decays,
the isocurvature perturbations between CDM (baryon) and radiation are given by
$S_{{\rm CDM}(b)\gamma} = 0 - (3/4)(\delta\rho_{\gamma}/\rho_{\gamma}) \neq 0$, 
neglecting inflaton fluctuations.
Moreover, the isocuravature perturbations produced in this way are anti-correlated 
with curvature perturbations $\zeta$ as $\zeta = -S_{{\rm CDM} (b)\gamma}/3$.
On the other hand, we have positively correlated isocurvature perturbations 
if CDM (baryon) is produced from the curvaton decay. 
In either case, since the isocurvature fluctuations are stringently constrained by 
CMB observations, they cause a serious cosmological difficulty in curvaton models. 

An obvious solution to the isocurvature problem is to produce 
both CDM and baryon number thermally after the curvaton decays.
Another interesting possibility is that CDM is produced from the inflaton 
and the baryon number is generated from the curvaton decay, or vice versa.
Then, the total matter isocurvature perturbations $S_m$ are written as
\begin{equation}
   S_m = \frac{\Omega_{\rm CDM}}{\Omega_m} S_{{\rm CDM}\gamma} 
   + \frac{\Omega_{b}}{\Omega_m} S_{b \gamma},
\end{equation}
where $\Omega_{\rm CDM}$, $\Omega_b$ and $\Omega_m$ are the density parameters 
of CDM, baryon and
matter ($\Omega_m = \Omega_{\rm CDM}+\Omega_b$), respectively. 
In this case, the contributions from CDM and 
baryon can cancel each other and hence
we can avoid the stringent constraint from the CMB because the temperature 
anisotropies are produced only through $S_m$ ( and $\zeta$).
When the cancellation occurs and $S_m$ is significantly reduced 
in comparison with $S_{{\rm CDM}}$ and $S_{b}$, 
it is said that CDM and baryonic 
isocurvature perturbations are ``compensated'' with each other~\cite{Holder:2009gd,Gordon:2009wx,Kawasaki:2011ze,Grin:2011tf,Grin:2013uya}. 

Furthermore, since the isocurvature perturbations can be anti-correlated with   
the adiabatic ones, temperature anisotropies on large scales can be 
reduced. 
In fact, the large scale temperature anisotropies by the Sachs-Wolfe effect
is approximately given by~\cite{Contaldi:2014zua} 
\begin{equation}
   \left\langle \left(\frac{\Delta T}{T} \right)^2\right\rangle = 
   \frac{1}{25}\left[ {\cal P}_{\zeta} + 4{\cal P}_{S_m}
   + 4{\cal P}_{\zeta S_m} + \frac{5}{6}{\cal P}_T\right],
   \label{eq:temp_aniso}
\end{equation}
where ${\cal P}_{\zeta}$ and ${\cal P}_{S_m}$ are the power spectra of
curvature and isocurvature perturbations, and ${\cal P}_{\zeta S_m}$
is the cross power spectrum of them. 
Here we have included the contribution from tensor perturbations on large scales
and ${\cal P}_T$ is their power spectrum.
From eq.~(\ref{eq:temp_aniso}), it is seen that 
$\Delta T/T$ decreases compared with a pure adiabatic case
if ${\cal P}_{S_m}+ {\cal P}_{\zeta S_m} < 0$.
Moreover, when the tensor mode exits, the isocurvature perturbations can 
compensate for its effect on $\Delta T/T$.
This effect may solve the tension between 
the Planck observation of the temperature fluctuations~\cite{Ade:2013uln} 
and the BICEP2 detection of the B-mode polarization~\cite{Ade:2014xna}
as pointed out in~\cite{Kawasaki:2014lqa}.
Although it is premature to conclude the B-mode detection, taking into account 
uncertainty of the foreground dust emission~\cite{Mortonson:2014bja,Flauger:2014qra}, 
need for reducing temperature fluctuations on large scales is also suggested 
from the analysis by~\cite{Smith:2014kka}, 
which shows that a negative tensor-to-scalar ratio 
$r \simeq -0.2$ gives a better fit to the Planck data.
The compensated isocurvature perturbations can not only avoid 
the stringent CMB constraint but also have a possibility to improve  
agreement with the observational data.  

In this paper we study a curvaton model where both CDM and 
baryonic isocurvature perturbations are produced.
We identify the curvaton as a right-handed sneutrino
in supersymmetric (SUSY) theories.  
In this model baryon asymmetry is generated via non-thermal leptogenesis
by decay of sneutrinos into left-handed (s)leptons and 
higgs(inos). 
On the other hand, gravitinos produced during reheating after inflation 
can be dark matter if they are the lightest SUSY particles (LSPs).
If not, LSPs produced by the gravitino decay account for dark matter. 
Then, baryon and CDM($=$LSPs) have correlated and anti-correlated 
isocurvature perturbations, respectively, and
they compensate for each other by taking appropriate model parameters.
The possibility of such compensated isocurvature perturbations in a sneutrino
curvaton model was pointed out in~\cite{Harigaya:2012up} but a detailed
analysis has not been performed.
We estimate the baryon number and the dark matter density as well as 
amplitudes of adiabatic and isocurvature perturbations in the sneutrino curvaton model. 
It is found that the compensated isocurvature perturbations are realized 
for the reheating temperature of $O(10^{9\mathchar`-10})$~GeV 
and the LSP mass of $O(0.1\mathchar`-1)$~TeV. 

This paper is organized as follows.
In Sec.~\ref{sec:compensate}, 
we briefly explain how the CDM/baryon isocurvature perturbations 
could compensate for each other in the curvaton model. 
Then, we derive conditions for 
the CDM/baryon isocurvature perturbations to cancel each other. 
We also show conditions for the isocurvature perturbations to compensate for 
the contribution of the tensor mode to temperature anisotropies. 
In Sec.~\ref{sec:model}, 
we investigate a sneutrino curvaton model 
where the curvaton is identified as the right-handed sneutrino. 
We estimate perturbations generated by the curvaton, the baryon asymmetry, and the CDM abundance in our model. 
In Sec.~\ref{sec:construction}, 
we express the above conditions for compensation with respect to model parameters of the curvaton scenario, 
such as the curvaton field value, the curvaton decay temperature and the mass of the LSP.
Finally, in Sec.~\ref{sec:conclusion}, we summarize our results.

\section{Compensated isocurvature perturbations}
\label{sec:compensate}

In this section, we briefly review 
the curvaton scenario~\cite{Enqvist:2001zp,Lyth:2001nq,Moroi:2001ct} 
and show how the isocurvature perturbations are generated. 
We explain how the compensated isocurvature perturbations could be realized in curvaton models.

In single-field inflation models, 
only the inflaton is responsible for the curvature perturbations. 
In curvaton scenarios, on the other hand, 
another light field called a curvaton is also the source of the adiabatic perturbations. 
During inflation, the curvaton is a subdominant component of the Universe. 
It acquires quantum fluctuations as the inflaton field does. 
After inflation ends, the inflaton begins its oscillation.  
Then, the curvaton field also starts to oscillate 
when the Hubble parameter becomes comparable to the curvaton mass. 
After that, the Universe is reheated via inflaton decay 
and a radiation dominated era is realized 
\footnote{
Here and hereafter, we assume that the curvaton starts to oscillate before the reheating 
since we identify the curvaton as the right-handed sneutrino 
whose mass is about $O(10^{12})$~GeV,
which is heavy enough to begin oscillation before the reheating 
for typical reheating temperatures.
}. 
During the radiation dominated period, 
the ratio of the curvaton density to the radiation energy density produced from
the inflaton decay increases in proportional to $a$, where $a$ is a scale factor.
Thus, the curvaton becomes a non-nigligible component of the Universe 
and the adiabatic perturbations evolve during this epoch. 
The evolution of the adiabatic perturbations stops when the curvaton decays 
and thereafter the curvature perturbations are conserved on super-horizon scales. 
The adiabatic perturbations in the curvaton scenario are given by 
\begin{equation}
   \zeta=\zeta_{{\rm inf}}+\frac{f_{{\rm dec}}}{3}S_{\sigma},
\end{equation}
where $\zeta$ is the curvature perturbation on the uniform density slicing, 
$\zeta_{{\rm inf}}$ is the curvature perturbation induced from the inflaton,
and $S_{\sigma}$ is the curvaton isocurvature perturbation 
which is given by $S_{\sigma}=3(\zeta_{\sigma}-\zeta_{{\rm inf}})$ 
with $\zeta_{\sigma}$ being the curvature perturbation 
on the uniform density slicing of the curvaton.
The parameter $f_{\rm dec}$ is defined by the energy density of the radiation produced by the inflation $\rho_r$ and that of the curvaton $\rho_\sigma$ as
\begin{eqnarray}
   f_{{\rm dec}}\equiv 
   \frac{3\rho_{\sigma}}{(4\rho_{r}+3\rho_{\sigma})} \Big|_{{\rm dec}}, 
\end{eqnarray}
where the subscript ``dec" denotes the value when the curvaton decays.

The curvaton scenario also generates matter isocurvature perturbations 
besides the adiabatic perturbations. 
As described in Sec.~\ref{sec:intro}, the matter isocurvature perturbations consist of 
the CDM and the baryon isocurvature perturbations. 
If the CDM (baryon number) is produced from the decay products of the inflaton 
and decouples from thermal bath before the curvaton decays, 
the residual isocurvature perturbations are given by 
\begin{eqnarray}
   S_{{\rm CDM}(b)\gamma} & \equiv & 3(\zeta_{{\rm CDM}(b)}-\zeta) \nonumber \\
   & = & -f_{{\rm dec}}S_{\sigma},
\end{eqnarray}
with $\zeta_{{\rm CDM}(b)}=\zeta_{{\rm inf}}$. 
If the baryon number (CDM) is produced by out-of-equillibrium decay of the curvaton, 
the residual isocurvature perturbations are given  by 
\begin{equation}
   S_{b({\rm CDM})\gamma}=(1-f_{{\rm dec}})S_{\sigma},
\end{equation}
with $\zeta_{b({\rm CDM})}=\zeta_{\sigma}$. 
In this paper, we study the case 
where CDM is produced from decay products of the inflaton  
and the baryon number is produced by out-of-equillibrium decay of the curvaton. 
In this case, 
the adiabatic perturbations and the total matter isocurvature perturbations 
are given by 
\begin{equation}
   \begin{pmatrix}
      \zeta \\[1.5ex]
      S_{m} \\
   \end{pmatrix}
   =
   \begin{pmatrix}
      1 & \dfrac{f_{{\rm dec}}}{3} \\[2.0ex]
      0 & {\cal T}_{S_mS_{\sigma}} \\
   \end{pmatrix}
   \begin{pmatrix}
      \zeta_{{\rm inf}} \\[1.5ex]
      S_{\sigma}
   \end{pmatrix},
   \label{eq:matrix}
\end{equation}
where the transfer function ${\cal T}_{S_mS_{\sigma}}$ is given by 
\begin{eqnarray}
   {\cal T}_{S_mS_{\sigma}} &=& -\frac{\Omega_{{\rm CDM}}}{\Omega_m} f_{{\rm dec}}
   + \frac{\Omega_b}{\Omega_m}(1 - f_{{\rm dec}}) \nonumber \\
   &=& \frac{\Omega_{b}}{\Omega_{m}}-f_{{\rm dec}}.
   \label{eq:transfer}
\end{eqnarray}
Then, one finds that 
${\cal P}_{\zeta}$, ${\cal P}_{S_m}$ and ${\cal P}_{\zeta S_m}$ 
are given by 
\begin{eqnarray}
   {\cal P}_{\zeta} &=& {\cal P}_{\zeta_{{\rm inf}}} + 
   \frac{f_{{\rm dec}}^2}{9}{\cal P}_{S_{\sigma}} \equiv (1 + R){\cal P}_{\zeta_{{\rm inf}}} 
   \label{eq:adiabatic_perturbation}, \\
   {\cal P}_{S_m} &=& {\cal T}_{S_mS_{\sigma}}^2 {\cal P}_{S_{\sigma}}, 
   \label{eq:isocurvature_perturbation} \\
   {\cal P}_{\zeta S_m} &=& 
   \frac{f_{{\rm dec}}}{3} {\cal T}_{S_mS_{\sigma}} {\cal P}_{S_{\sigma}},
   \label{eq:cross}
\end{eqnarray}
where $R$ is defined as the ratio of the power spectra of the curvature perturbations 
from the inflaton to that from the curvaton. 
It should be noted that
the cross power spectrum ${\cal P}_{\zeta S_m}$ is negative 
for ${\cal T}_{S_m S_{\sigma}}< 0$ and then anti-correlated isocurvature perturbations
can be realized. 
It is found that the CDM/baryon isocurvature perturbations cancel each other 
and the total matter isocurvature perturbations vanish 
for ${\cal T}_{S_m S_{\sigma}} = 0$, in other words, 
$f_{{\rm dec}} = \Omega_b/\Omega_m \simeq 0.16$ in eq.~(\ref{eq:transfer})
for values of density parameter obtained from Planck+WP+highL+BAO~\cite{Ade:2013zuv}.

Next let us focus on the temperature anisotropies. 
On large scales, the CMB anisotropies originate from the Sachs-Wolfe effect 
and are given by~\cite{Contaldi:2014zua} 
\begin{equation}
   \left( \frac{\Delta T}{T}\right)_{{\rm SW}} 
   \simeq -\frac{1}{5}\zeta - \frac{2}{5}S_m + \frac{1}{2}h_{ij}n^in^j,
\end{equation}
where $h_{ij}$'s are the tensor perturbations and $n$ is a unit vector. 
The correlation of the temperature anisotropies on large scales is approximately
given by eq.~(\ref{eq:temp_aniso}),
\begin{eqnarray}
   \left\langle \left(\frac{\Delta T}{T} \right)^2\right\rangle
   &=& \frac{1}{25}\left[ {\cal P}_{\zeta} + 4{\cal P}_{S_m} + 4{\cal P}_{\zeta S_m} 
   + \frac{5}{6}{\cal P}_T \right] \nonumber \\
   &=& \frac{1}{25}{\cal P}_{\zeta}\left[ 1 + 4 B_m^2+ 
   4 B_m \cos\theta_m + \frac{5}{6}r \right],
   \label{eq:temp_aniso2}
\end{eqnarray}
where $r\equiv{\cal P}_T/{\cal P}_{\zeta}$. 
Here we have defined $B_m$ and $\cos \theta_m$ as 
\begin{eqnarray}
\label{eq:B}
   B_m &\equiv& \sqrt{\frac{{\cal P}_{S_m}}{{\cal P}_{\zeta}}},\\
\label{eq:cos}
 \cos \theta_m &\equiv& \frac{{\cal P}_{\zeta S_m}}{\sqrt{{\cal P}_{\zeta}{\cal P}_{S_m}}}.
\end{eqnarray}
If the total matter isocurvature perturbations are anti-correlated 
with the adiabatic perturbations, 
the third term of eq.~(\ref{eq:temp_aniso2}) is negative. 
In that case, it is possible that 
the isocurvature perturbations compensate for 
the tensor contributions to the temperature anisotropies on large scales~\cite{Kawasaki:2014lqa}. 
In other words, 
\begin{eqnarray}
   4{\cal P}_{S_m} + 4{\cal P}_{\zeta S_m} + \frac{5}{6}{\cal P}_T = 0,
   \label{eq:compensate}
\end{eqnarray}
could be satisfied.
From eqs.~(\ref{eq:transfer}), (\ref{eq:isocurvature_perturbation}) and (\ref{eq:cross}),
we can rewrite it as follows,
\begin{eqnarray}
   4 {\cal P}_{S_{\sigma}}\left( \frac{\Omega_b}{\Omega_m}-f_{{\rm dec}}\right)
   \left( \frac{\Omega_b}{\Omega_m}-\frac{2}{3}f_{{\rm dec}}\right)
   +\frac{5}{6}{\cal P}_T=0.
   \label{eq:compensate2}
\end{eqnarray}

The required isocurvature perturbations to satisfy eq.~(\ref{eq:compensate2}) 
are realized by taking appropriate parameters.
\begin{figure}[tb]
\begin{center}
\includegraphics[width=120mm]{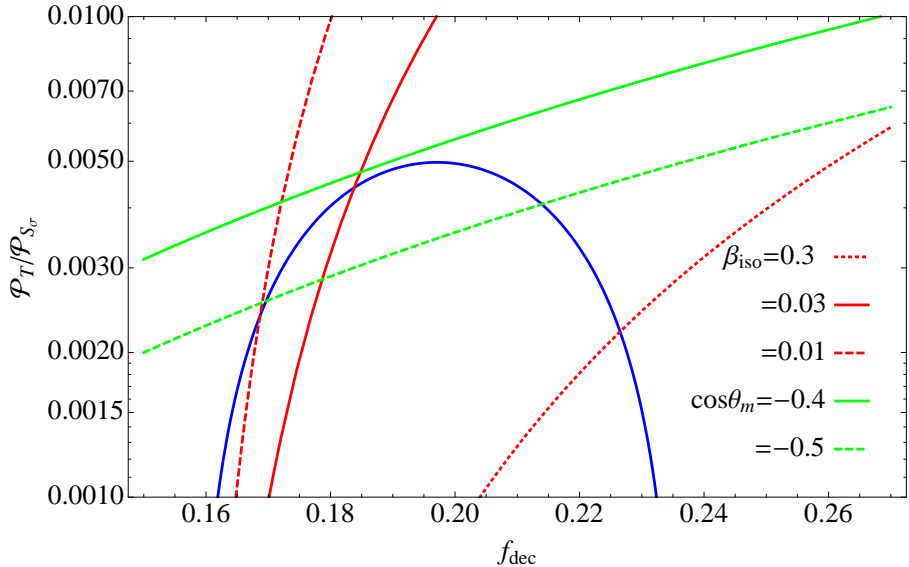}
\end{center}
\caption{
The solution for eq.~(\ref{eq:compensate2}) 
in $f_{\rm dec}$-${\cal P}_T/{\cal P}_{S_{\sigma}}$ plane.
The red lines show contours of isocurvature fractions 
$\beta_{{\rm iso}}\equiv {\cal P}_{S_m}/({\cal P}_{\zeta}+{\cal P}_{S_m})$
and the green lines show those of $\cos \theta_m$ for $r=0.2$.
}
\label{fig:Tensor_f}
\end{figure}
In Fig.~\ref{fig:Tensor_f}, we plot the solution of eq.~(\ref{eq:compensate2})
with the blue solid line
in $f_{\rm dec}$-${\cal P}_T/{\cal P}_{S_{\sigma}}$ plane.
We also plot contours of isocurvature fractions $\beta_{{\rm iso}}$,
defined as $\beta_{{\rm iso}}\equiv{\cal P}_{S_m}/({\cal P}_{\zeta}+{\cal P}_{S_m})$,
and $\cos\theta_m$ defined in eq.~(\ref{eq:cos}),
where we assume the tensor-to-scalar ratio $r=0.2$.
From observations of Planck and BICEP2 collaboration,
the allowed parameter regions are $\beta_{{\rm iso}}\lesssim 0.03$ and 
$\cos \theta_m\lesssim -0.4$~\cite{Kawasaki:2014fwa}.
One can see that the solution exists around $f_{\rm dec} \simeq 0.18$ and
${\cal P}_T/{\cal P}_{S_{\sigma}} \sim O(10^{-3})$. 
Hence, when we choose the parameters ${\cal P}_T/{\cal P}_{S_{\sigma}}$
and $f_{\rm dec}$ appropriately,
we can compensate for the contribution from the tensor mode in the CMB temperature
anisotropies on large scales, due to the anti-correlated isocurvature perturbations.

In this paper, 
we investigate following two cases: 
\begin{itemize}
   \item The CDM/baryon isocurvature perturbations cancel each other ($S_m = 0$).
   \item The isocurvature perturbations compensate for the tensor contribution 
   to the temperature anisotropies. 
\end{itemize}
As explained above, the former case is realized for $f_{{\rm dec}} \simeq 0.16$ and 
the latter case is realized 
when the condition given by eq.~(\ref{eq:compensate2}) is satisfied. 

\section{Sneutrino curvaton scenario}
\label{sec:model}

In this section, we focus on a curvaton scenario based on supersymmetric theories. 
We identify the curvaton 
with a right-handed sneutrino in supersymmetric theories.
We introduce three generations of right-handed neutrino chiral multiplets $N_i$ 
with masses $M_i$ ($i=1,2,3$)
into the minimal supersymmetric standard model 
in order to explain neutrino masses by the seesaw mechanism~\cite{Yanagida:1979as}. 
A superpotential of the right-handed neutrinos is given by
\begin{equation}
   W = \frac{1}{2}M_iN_iN_i + \lambda_{ij} N_i L_j H_u,
\end{equation}
where $L_j$ and $H_u$ are chiral multiplets of the lepton doublets 
and the up-type Higgs, respectively. 
We assume mass hierarchy as $M_1\ll M_2< M_3$ and that
the lightest right-handed sneutrino $\widetilde{N}_1$ 
plays the role of the curvaton
\footnote{
$\widetilde{N}_2$ or $\widetilde{N}_3$ can be the inflaton which causes chaotic inflation~\cite{Murayama:1992ua,Murayama:2014saa}.
}.

We assume that the baryon asymmetry is generated via decay of the sneutrino curvaton.
We also assume that gravitinos produced during reheating are responsible for the observed CDM abundance.
In this way,
baryon and CDM have correlated and anti-correlated 
isocurvature perturbations, respectively, and
they compensate for each other by taking appropriate model parameters.

In the following, we express 
the power spectra and $f_{{\rm dec}}$
which are discussed in the previous section,
in terms of model parameters of the sneutrino curvaton scenario.
We also estimate the baryon asymmetry and the CDM abundance. 

\subsection{Power spectra and $f_{{\rm dec}}$}
\label{sec:conditions}

We first express the power spectra
in terms of model parameters. 
Hereafter, we assume that the curvaton has a quadratic potential, 
$V(\sigma) = \frac{1}{2}M_1^2\sigma^2$. 
The power spectrum of curvature perturbations generated from the inflaton fluctuations is given by 
\begin{equation}
   {\cal P}_{\zeta_{{\rm inf}}}(k) = \frac{H_*^2(k)}{8\pi^2M_{{\rm pl}}^2\epsilon_*(k)},
   \label{eq:zeta_inf}
\end{equation}
where $H$ is the Hubble parameter with the star denoting the epoch of horizon exit, 
$k = a_*H_*$, 
$M_{{\rm pl}}$ is the reduced Planck mass 
and $\epsilon_*$ is a slow-roll parameter defined as 
$\epsilon_* \equiv \frac{M_{{\rm pl}}^2}{2}(\frac{V_{\phi}}{V})^2|_{\phi_*}$ 
where the subscript $\phi$ denotes $\partial / \partial \phi$. 
The power spectrum of the curvaton isocurvature perturbations is 
given by~\cite{Lyth:2002my} 
\begin{equation}
   {\cal P}_{S_{\sigma}} = \frac{4}{\sigma_*^2}\left( \frac{H_*}{2\pi} \right)^2,
\end{equation}
where $\sigma_*$ denotes the curvaton field value at the horizon exit. 
The power spectrum of the tensor perturbations is given by
\begin{eqnarray}
   {\cal P}_T=\frac{8}{M_{{\rm pl}}^2}\left( \frac{H_*}{2\pi} \right)^2.
\end{eqnarray}
The ratio of the power spectra of the tensor to curvaton isocurvature perturbations 
is therefore given by 
\begin{equation}
   \frac{{\cal P}_T}{{\cal P}_{S_{\sigma}}} 
   = 2\left( \frac{\sigma_*}{M_{{\rm pl}}}\right)^2. 
   \label{eq:P/P}
\end{equation}
One finds that ${\cal P}_T/{\cal P}_{S_{\sigma}}$ 
depends only on the curvaton field value $\sigma_*$. 

Next we estimate
the ratio of the curvaton to the radiation energy density 
at the curvaton decay time, $\rho_{\sigma}/\rho_r|_{{\rm dec}}$, which appears in the parameter $f_{\rm dec}$. 
Since the ratio of the energy densities does not change 
while both the inflaton and the curvaton are oscillating, 
it is given by 
\begin{equation}
   \frac{\rho_{\sigma}}{\rho_r}\Big|_{{\rm dec}} = \frac{\rho_{\sigma}}{\rho_\phi}\Big|_{{\rm osc}}
   \left( \frac{a_{{\rm dec}}}{a_{{\rm reh}}}\right),
   \label{eq:ratio}
\end{equation}
where subscripts ``${{\rm osc}}$", ``${{\rm dec}}$" and ``${{\rm reh}}$" 
respectively denote the time of the onset of the curvaton oscillation, 
the curvaton decay  and the reheating. 
We take 
$H_{{\rm osc}} =M_1$, $H_{{\rm dec}} = \Gamma_{\sigma}$ and 
$H_{{\rm reh}} = \Gamma_{\phi}$,
where $\Gamma_{\sigma}$ and $\Gamma_{\phi}$ are 
decay rates of the curvaton and the inflaton. 
Then, we define the reheating (curvaton decay) temperature as 
$T_{{\rm reh}({\rm dec})}\equiv 
\left(\frac{90}{\pi^2 g_*}M_{{\rm pl}}^2\Gamma_{\phi(\sigma)}^2\right)^{1/4}$. 
Assuming that the curvaton field stays at $\sigma_*$ until its oscillation, 
$\rho_{\sigma}|_{{\rm osc}} = \frac{1}{2}M_1^2\sigma_*^2$. 
We also assume that 
the inflaton dominates the energy density of the Universe
when the curvaton starts to oscillate.
Then, $\rho_{\sigma}/\rho_{\phi}|_{{\rm osc}}$ is written as 
\begin{eqnarray}
   \frac{\rho_{\sigma}}{\rho_\phi}\Big|_{{\rm osc}} & = & 
   \frac{\frac{1}{2}M_1^2\sigma_*^2}{3M_{{\rm pl}}^2H_{{\rm osc}}^2} \nonumber \\
   & = & \frac{M_1^2\sigma_*^2}{6M_{{\rm pl}}^2M_1^2} = 
   \frac{1}{6}\left( \frac{\sigma_*}{M_{{\rm pl}}}\right)^2.
\end{eqnarray}
In the second line, we have used $H_{{\rm osc}} = M_1$. 
Since the temperature of the Universe is inversely proportional to the scale factor, 
eq.~(\ref{eq:ratio}) is written as 
\begin{equation}
   \frac{\rho_{\sigma}}{\rho_r} \Big|_{{\rm dec}} 
   = \frac{1}{6} \left( \frac{\sigma_*}{M_{{\rm pl}}}\right)^2\frac{T_{\rm reh}}{T_{{\rm dec}}}. 
   \label{eq:f_dec}
\end{equation}
Since the parameter $f_{{\rm dec}}$ 
is determined by $\rho_{\sigma}/\rho_{r}|_{{\rm dec}}$, 
it depends on the curvaton field value $\sigma_*$ 
and the ratio of the temperature $T_{\rm reh}/T_{{\rm dec}}$. 

Finally, we also estimate the parameter $r$ 
defined in eq.~(\ref{eq:temp_aniso2}).
The tensor-to-scalar ratio $r$ is given by
\begin{eqnarray}
   r &=& \frac{{\cal P}_T}{(1+R){\cal P}_{\zeta_{{\rm inf}}}} \nonumber \\
   &=& 16\epsilon_*\left[ 1 + \frac{8}{9}\epsilon_*f_{{\rm dec}}^2
   \left( \frac{M_{{\rm pl}}}{\sigma_*}\right)^2\right]^{-1}. 
   \label{eq:r}
\end{eqnarray}
From eqs.~(\ref{eq:P/P}), (\ref{eq:f_dec}) and (\ref{eq:r}),
it is found that ${\cal P}_T/{\cal P}_{S_{\sigma}}$ 
depends only on $\sigma_*$,
$f_{{\rm dec}}$ depends on $\sigma_*$ and $T_{\rm reh}/T_{{\rm dec}}$, and 
$r$ depends on $\sigma_*$, $T_{\rm reh}/T_{{\rm dec}}$ and $\epsilon_*$.

\subsection{Baryon asymmetry}
\label{sec:baryon}

$\widetilde{N}_1$ non-thermally decays into (s)leptons and higgs(inos) 
as well as their anti-particles. 
Since CP symmetry is generally violated in the interaction of leptons, 
lepton asymmetry is generated by the decay of $\widetilde{N}_1$. 
The lepton asymmetry is partially converted into the baryon asymmetry 
through the sphaleron process~\cite{Klinkhamer:1984di}.

The relation between the baryon asymmetry and the lepton asymmetry in supersymmetic theories 
is given by~\cite{Khlebnikov:1988sr,Harvey:1990qw}
\begin{equation}
   \frac{n_B}{s} = - \frac{8}{23}\frac{n_L}{s},
   \label{eq:baryon_asym}
\end{equation}
where $s$, $n_B$ and $n_L$ are respectively the entropy density, 
the baryon number density and the lepton number density. 
The lepton asymmetry generated by the decay of $\widetilde{N}_1$ is given by 
\begin{equation}
   \frac{n_L}{s} = \epsilon_1 \frac{n_{\widetilde{N}_1}}{s}\Big|_{{\rm dec}},
   \label{eq:lepton_asym}
\end{equation}
where the parameter $\epsilon_1$ denotes CP asymmetry by the decay of $\widetilde{N}_1$,
which is defined as 
\begin{equation}
   \epsilon_1 \equiv
   \frac{\Gamma (\widetilde{N}_1 \to \widetilde{L} + H_u) 
   - \Gamma (\widetilde{N}_1 \to \widetilde{L}^* + H_u^*)}
   {\Gamma (\widetilde{N}_1 \to \widetilde{L} + H_u) 
   + \Gamma (\widetilde{N}_1 \to \widetilde{L}^* + H_u^*)} ,
\end{equation}
where $\Gamma$'s are decay rates 
and tildes denote superpartners of the right-handed neutrinos and lepton doublets. 
Considering the interference of the one-loop diagrams 
with the tree level coupling in supersymmetric theories, 
the parameter $\epsilon_1$ is given by~\cite{Covi:1996wh}
\begin{equation}
   \epsilon_1 = - \frac{1}{8\pi}\frac{1}{(\lambda \lambda^{\dagger})_{11}} 
   \sum_{i = 2,3} {\rm Im}\left[ \left\{ (\lambda \lambda^{\dagger})_{1i}\right\}^2\right] 
   f\left( \frac{M_i^2}{M_1^2}\right), 
\end{equation}
where the function $f(x)$ is defined as 
\begin{equation}
   f(x) \equiv \sqrt{x} \ln \left( 1+\frac{1}{x}\right) + \frac{2\sqrt{x}}{x-1}. 
\end{equation}
Using the mass hierarchy $M_1\ll M_2< M_3$ and the seesaw relation,
$\epsilon_1$ is estimated as 
\begin{equation}
   \epsilon_1 \simeq 
   \frac{3}{8\pi}\frac{M_1}{\langle H_u\rangle^2}m_{\nu3}\delta_{{\rm eff}},
\end{equation}
where $\langle H_u\rangle$ denotes the VEV of the up-type Higgs bosons, 
$m_{\nu3}$ is the heaviest neutrino mass produced via the seesaw mechanism 
and $\delta_{\rm eff}$ is a CP violating phase with $|\delta_{{\rm eff}}|\leq 1$. 

The number density of $\widetilde{N}_1$ at its decay is given in terms of $\sigma_*$ as 
\begin{eqnarray}
   n_{\widetilde{N}_1}|_{{\rm dec}} 
   &=& \frac{1}{2}M_1\sigma_*^2
   \left( \frac{a_{{\rm osc}}}{a_{{\rm reh}}}\right)^3 
   \left( \frac{a_{{\rm reh}}}{a_{{\rm dec}}}\right)^3 \nonumber \\
   &=& \frac{1}{2}M_1\sigma_*^2
   \left( \frac{\Gamma_{\phi}}{M_1}\right)^2
   \left( \frac{T_{{\rm dec}}}{T_{\rm reh}}\right)^3. 
\end{eqnarray}
In the second line, we have used the following relation, 
\begin{equation}
   \left( \frac{a_{{\rm osc}}}{a_{{\rm reh}}}\right)^3
   =\left( \frac{H_{{\rm reh}}}{H_{{\rm osc}}}\right)^2
   = \left( \frac{\Gamma_{\phi}}{M_1}\right)^2. 
\end{equation}
By using the relations, $3M_{{\rm pl}}^2\Gamma_{\phi}^2 
= \frac{\pi^2}{30}g_*T_{\rm reh}^4$ 
and $s|_{{\rm dec}}= \frac{2\pi^2}{45}g_{*s}T_{{\rm dec}}^3$, 
we obtain 
\begin{equation}
   \frac{n_{\widetilde{N}_1}}{s}\Big|_{{\rm dec}} 
  = \frac{1}{8}\left( \frac{\sigma_*}{M_{{\rm pl}}}\right)^2\frac{T_{\rm reh}}{M_1}. 
\end{equation}

From eqs.~(\ref{eq:baryon_asym}) and (\ref{eq:lepton_asym}), 
the generated baryon asymmetry by the decay of the sneutrino curvaton is given by 
\begin{eqnarray}
   \frac{n_B}{s} &=& - \frac{1}{23}\epsilon_1
   \left( \frac{\sigma_*}{M_{{\rm pl}}}\right)^2\frac{T_{\rm reh}}{M_1} \nonumber \\
   &\simeq& - 1.5\times10^{-11}\left( \frac{\sigma_*}{10^{17}\,{\rm GeV}}\right)^2 
   \left( \frac{T_{\rm reh}}{10^9\,{\rm GeV}}\right)
   \left\{ \left( \frac{174\,{\rm GeV}}{\langle H_u\rangle}\right)^2 
   \left( \frac{m_{\nu3}}{0.05\,{\rm eV}}\right)\delta_{{\rm eff}} \right\}. 
   \label{eq:baryon}
\end{eqnarray}
Hereafter, we take $\langle H_u \rangle \simeq 174\,{\rm GeV}$
assuming $\langle H_u \rangle$ is larger than 
the VEV of the down-type Higgs, $\langle H_d \rangle$. 
As for the heaviest neutrino mass and the CP violating phase, 
we take $m_{\nu3} = 0.05 \, {{\rm eV}}$ and $\delta_{{\rm eff}} = -1$. 
We treat $\sigma_*$ and $T_{\rm reh}$ as free parameters and 
choose them in order to obtain the present baryon asymmetry, 
$n_B/s \simeq 8.7\times10^{-11}$~\cite{Beringer:1900zz}.

\subsection{CDM abundance}
\label{sec:cdm}

In supersymmetric theories, 
gravitinos are copiously produced during reheating. 
The gravitinos could be dark matter 
if they are the lightest SUSY particles (LSPs). 
If not, decay products of the gravitinos may account for dark matter. 
In the following,
we assume that gravitinos produced during reheating 
is the dominant source of the dark mater abundance
and that non-thermal production of gravitinos by decay of scalar condensations 
and thermal production of dark matter are negligible.

The gravitino abundance for the reheating temperature of $O(10^9)$~GeV 
is approximately given by~\cite{Kawasaki:2008qe}
\begin{equation}
   Y_{3/2} \equiv \frac{n_{3/2}}{s} \simeq 1.4\times10^{-13}
   \left[1 + 0.6\left( \frac{m_{1/2}}{m_{3/2}}\right)^2 \right]
   \left( \frac{T_{\rm reh}}{10^9\,{\rm GeV}}\right),
   \label{eq:gravitino}
\end{equation}
where $n_{3/2}$, $m_{3/2}$ and $m_{1/2}$ respectively denote 
the gravitino number density, the gravitino mass 
and the unified gaugino mass at the GUT scale. 

From eq.~(\ref{eq:gravitino}), the dark matter density parameter is given by 
\begin{equation}
   \Omega_{{\rm CDM}}h^2 \simeq 3.8\times10^{-2}
   \left[ 1 + 0.6\left( \frac{m_{1/2}}{m_{3/2}}\right)^2\right]
   \left( \frac{m_{{\rm LSP}}}{1\,{\rm TeV}}\right)
   \left( \frac{T_{\rm reh}}{10^9\,{\rm GeV}}\right).
   \label{eq:CDM_gravitino}
\end{equation}
In the case where the gravitino is not the LSP,
assuming $m_{3/2}$ is much heavier than $m_{1/2}$, 
the second term of eq.~(\ref{eq:gravitino}) is negligible. 

As mentioned in Sec.~\ref{sec:baryon},
we treat $T_{\rm reh}$ as a free parameter. 
The mass parameters, such as $m_{3/2}$, $m_{1/2}$ and $m_{{\rm LSP}}$, 
depend on SUSY breaking models. 
Therefore, we also treat the mass parameters as free parameters. 
We choose $T_{\rm reh}$ and the mass parameters 
so as to obtain the correct dark matter density parameter,
$\Omega_{{\rm CDM}}h^2 \simeq 0.12$~\cite{Ade:2013zuv}.

\section{Model parameters}
\label{sec:construction}

In the previous section, we have investigated 
the generation of the CDM/baryon isocurvature perturbations
and estimated the baryon asymmetry and the CDM abundance 
in the context of the sneutrino curvaton model. 
In this section, we investigate parameter regions in the model which realizes
the compensated isocurvature perturbations and the case
where the anti-correlated isocurvature perturbations compensate for the
tensor contribution in the large scale CMB temperature anisotropies.

\subsection{Compensation for isocurvature perturbations}
\label{sec:S_m=0}

We first consider the case where 
the compensation for the CDM/baryon isocurvature perturbations occurs, 
that is, $S_m = 0$,
which requires $f_{{\rm dec}} \simeq 0.16$. 
In this case, from eqs.~(\ref{eq:f_dec}) and (\ref{eq:baryon})
we find that the curvaton decay temperature $T_{\rm dec}$
is fixed to
\begin{eqnarray}
T_{{\rm dec}} \simeq 
7\times10^6\,{\rm GeV}.
\end{eqnarray}
This is a robust prediction in our model.

 $T_{\rm reh}$ and $\sigma_*$ depend on
 the mass parameters $m_{1/2}$, $m_{3/2}$ and $m_{\rm LSP}$.
In order to obtain the observed CDM density
parameter,
$T_{\rm reh}$ is given by 
\begin{equation}
   T_{\rm reh} \simeq 
   3\times 10^{9}\,{\rm GeV}
   \left[ 1 + 0.6\left(\frac{m_{1/2}}{m_{3/2}}\right)^2\right]^{-1}
   \left( \frac{m_{\rm LSP}}{1\,{\rm TeV}}\right)^{-1},
\end{equation}
from eq.~(\ref{eq:CDM_gravitino}).
From eq.~(\ref{eq:baryon}) and the above value of $T_{\rm reh}$,
the condition for successful baryogenesis leads to
\begin{equation}
   \sigma_* \simeq 
   1\times10^{17}\,{\rm GeV}
   \left[ 1 + 0.6\left(\frac{m_{1/2}}{m_{3/2}}\right)^2\right]^{1/2}
   \left( \frac{m_{\rm LSP}}{1\,{\rm TeV}}\right)^{1/2}.
\end{equation}

The curvaton scenario is known as a model which may generate
large local primordial non-Gaussianity.
The non-linearity parameter $f^{\rm loc}_{\rm NL}$ has been commonly used as
a parameter characterizing the local primordial non-Gaussianity.
Planck collaboration gives a tight constraint on $f^{\rm loc}_{\rm NL}$ 
as $f^{\rm loc}_{\rm NL} < O(10)$~\cite{Ade:2013ydc}.
Basically, in the case where the contribution from the curvaton fluctuations
is dominant in the power spectrum of the curvature perturbation, that is, $R \gg 1$,
the non-linearity parameter is estimated as $f^{\rm loc}_{\rm NL} \sim 1/f_{\rm dec}$.
On the other hand, in the case with $R \ll 1$, $f^{\rm loc}_{\rm NL}$ is suppressed 
as $f^{\rm loc}_{\rm NL} \sim R^2 \times 1/f_{\rm dec}$.
In our model,
$R$ is given by 
$R=\frac{8}{9}\epsilon_* f_{{\rm dec}}^2 \left(\frac{M_{{\rm pl}}}{\sigma_*} \right)^2$.
Therefore, the parameter region we consider corresponds to the case with $R \sim O(0.1)$, 
and hence our model predicts $f^{\rm loc}_{\rm NL} = O(0.1)$, which is consistent with the Planck result.

It is found that
the curvaton field value $\sigma_*$ is slightly smaller than $M_{{\rm pl}}$ and
the reheating temperature $T_{\rm reh}$ is of $O(10^{9\mathchar`-10})$~GeV,
where we assume $m_{\rm LSP}\sim m_{1/2}\sim O(1)$~TeV and $m_{3/2}\gtrsim O(1)$~TeV.
The reheating temperature of $O(10^{9\mathchar`-10})$ GeV is a typical one in chaotic inflation models~\cite{Kawasaki:2000ws,Harigaya:2014pqa,Harigaya:2014roa}.
If the gravitino is the LSP and much lighter than the weak scale,
$\sigma_*$ is far above $M_{\rm pl}$, which is incompatible with the curvaton scenario.
We stress that 
the curvaton decay temperature $T_{{\rm dec}}$ is uniquely predicted as $T_{{\rm dec}} \simeq7 \times 10^{6}$~GeV in this model.

\subsection{Compensation for tensor modes}
\label{sec:tensor}

\begin{figure}[htbp]
   \begin{minipage}{0.5\hsize}
      \begin{center}
      \includegraphics[width=70mm]{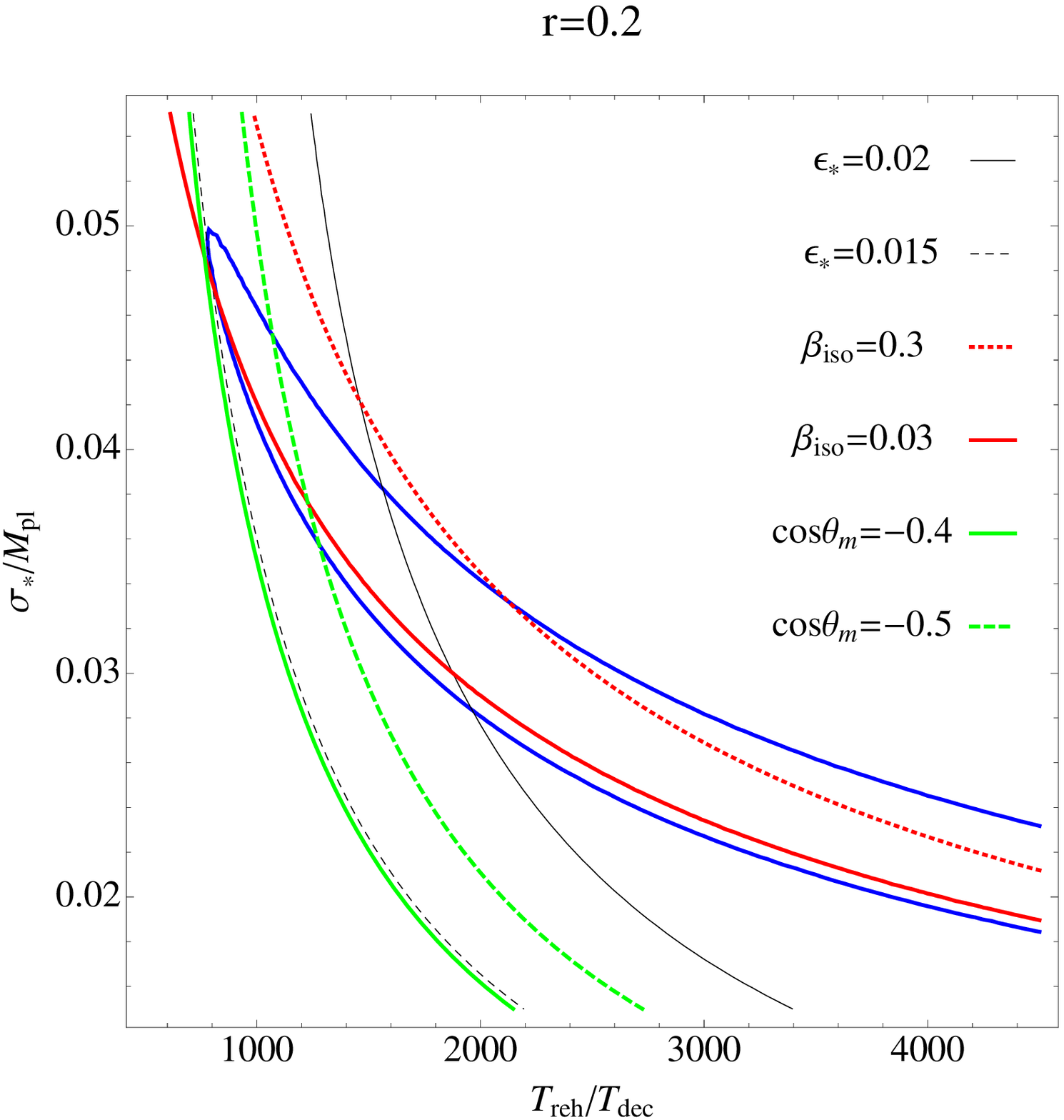}
      \end{center}
   \end{minipage}
   \begin{minipage}{0.5\hsize}
      \begin{center}
      \includegraphics[width=70mm]{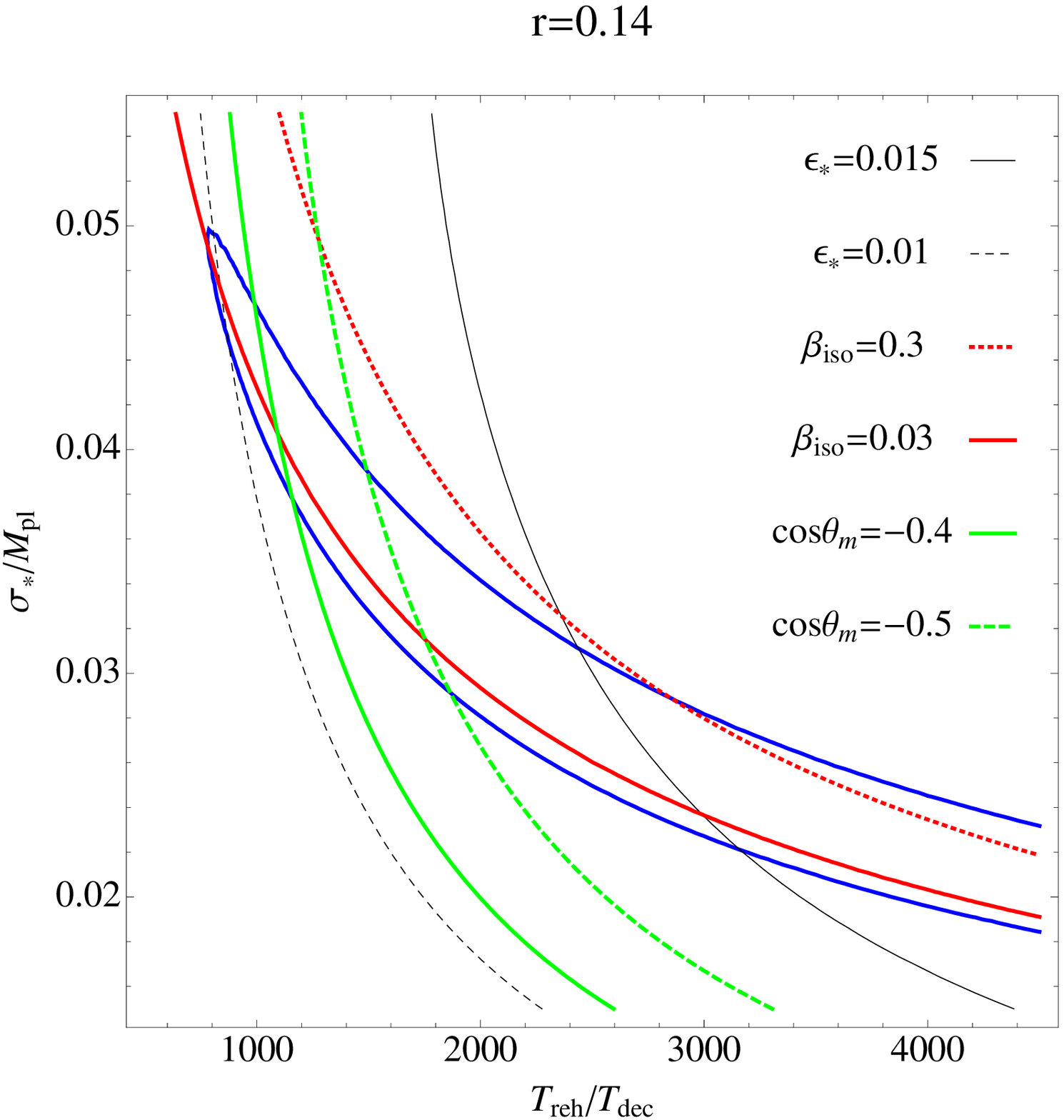}
      \end{center}
   \end{minipage}
   \caption{The solutions for eq.~(\ref{eq:compensate2}) 
   in $\sigma_*$-$T_{\rm reh} / T_{\rm dec}$ plane are denoted by the blue lines. 
   The black lines show the condition to obtain the tensor-to-scalar ratio $r$ 
   for each slow-roll parameter $\epsilon_*$.
   The red lines are contours of the isocurvature fractions $\beta_{{\rm iso}}$ 
   and the green lines are those of $\cos \theta_m$.
   The allowed parameter region, $\beta_{{\rm iso}}<0.03$ and $\cos\theta_m< -0.4$, 
   is below the red solid line and  above the green solid line.
   We show the cases for $r=0.2$ (left panel) and $r=0.14$ (right panel).}
   \label{fig:T_sigma}

\end{figure}

Next, we consider the case 
where the anti-correlated isocurvature perturbations compensate for the tensor modes. 
In Sec.~\ref{sec:compensate}, 
we have derived the condition, eq.~(\ref{eq:compensate2}), on 
${\cal P}_T/{\cal P}_{S_{\sigma}}$ and $f_{{\rm dec}}$
to realize the compensation for tensor modes.
In Sec.~\ref{sec:conditions}, 
we have shown the dependence of  ${\cal P}_T/{\cal P}_{S_{\sigma}}$ and $f_{{\rm dec}}$
on model parameters, $\sigma_*$ and $T_{{\rm reh}}/T_{{\rm dec}}$.
In Fig.~\ref{fig:T_sigma},
we show the conditions to realize the compensation for tensor modes with the blue solid line
in $\sigma_*$-$T_{\rm reh} / T_{\rm dec}$ plane,
corresponding to Fig.~\ref{fig:Tensor_f}.
Note that they have no dependence on the tensor-to-scalar ratio $r$.
The black lines show the condition to obtain $r$ 
for each slow-roll parameter $\epsilon_*$,
estimated by eq.~(\ref{eq:r}).
The red lines are contours of the isocurvature fractions
$\beta_{{\rm iso}}\equiv {\cal P}_{S_m}/({\cal P}_{\zeta} + {\cal P}_{S_m})$ and 
the green lines are those of $\cos\theta_m$.

As mentioned in Sec.~\ref{sec:compensate},
constraints on $\beta_{{\rm iso}}$ and $\cos\theta_m$ are given by
$\beta_{{\rm iso}} \lesssim 0.03$ and 
$\cos\theta_m\lesssim -0.4$~\cite{Kawasaki:2014fwa}.
Furthermore, 
since the observed curvature perturbations are nearly scale invariant
and the spectral index is $n_s\simeq 0.96$~\cite{Ade:2013uln},
the slow-roll parameter should be small as $\epsilon_* \lesssim O(0.01)$.
Therefore, we find that
\begin{eqnarray}
\label{eq:sigmaT}
\sigma_* / M_{\rm pl} \sim O(0.01)~~{\rm and}~~T_{\rm reh} / T_{\rm dec} \sim O(10^3),
\end{eqnarray}
are required
in order to realize the compensation for tensor modes.

\begin{figure}[htbp]
\begin{center}
\includegraphics[width=78mm]{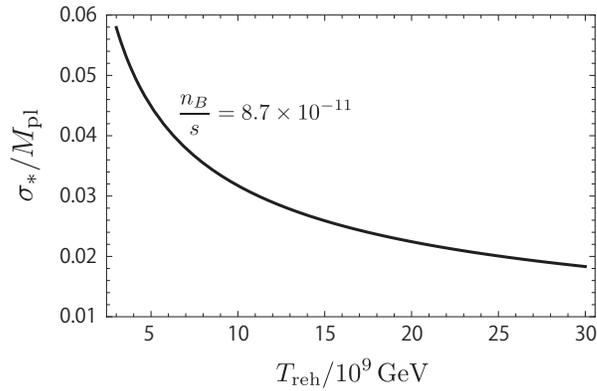}
\end{center}
\caption{
The present baryon asymmetry in $\sigma_*$-$T_{\rm reh}$ plane.}
\label{fig:sigma_TR}
\end{figure}

In Fig.~\ref{fig:sigma_TR},
we plot the relation between 
the curvaton field value $\sigma_*$ and the reheating temperature $T_{{\rm reh}}$ 
to obtain the present baryon asymmetry.
From eq.~(\ref{eq:sigmaT}) and Fig.~\ref{fig:sigma_TR},
we find that $T_{\rm reh} = O(10^9)$~GeV is required. 

Once the allowed  values of the reheating temperature $T_{\rm reh}$ and the curvaton field value $\sigma_*$
are confined,
the mass parameters $m_{3/2}$, $m_{1/2}$ and $m_{{\rm LSP}}$ 
are determined from eq.~(\ref{eq:CDM_gravitino}) to obtain the observed dark matter abundance.
We find that
\begin{eqnarray}
m_{3/2}\gtrsim O(0.1\mathchar`-1)~{\rm TeV},~~m_{1/2}\sim m_{\rm LSP}\sim O(0.1\mathchar`-1)~{\rm TeV},
\end{eqnarray}
are required if the gravitino is as heavy as or heavier than 
minimal supersymmetric standard model particles.
It is remarkable that required soft masses are consistent with TeV scale SUSY.
If the gravitino is the LSP and much lighter than the weak scale,
$m_{1/2}$ is required to be smaller,
which is inconsistent with various experiments.

\begin{figure}[tb]
\begin{center}
\includegraphics[width=78mm]{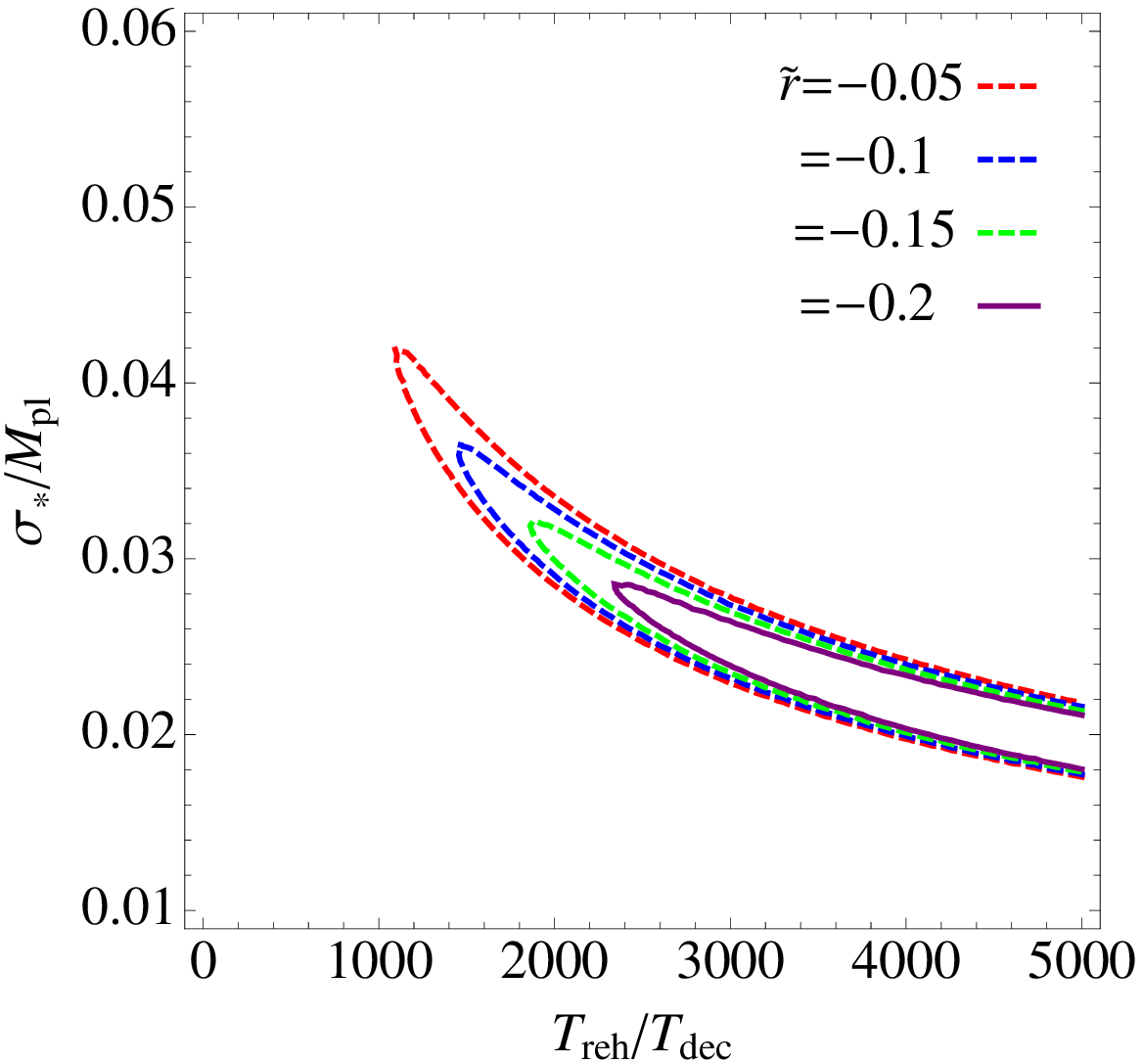}
\end{center}
\caption{
The solutions for eq.~(\ref{eq:negative_tensor}) in $\sigma_*$-$T_{\rm reh}$ plane.
We fix the slow-roll parameter $\epsilon_*=0.01$.}
\label{fig:negative_tensor}
\end{figure}

As we have mentioned in Sec.~\ref{sec:S_m=0},
the curvaton model might generate
large primordial non-Gaussianity, $f^{\rm loc}_{\rm NL}$.
The parameter region we consider, however, corresponds to the case with $R \sim O(0.1)$
and hence our model is consistent with the Planck result.

As we have shown,
our model realizes
the isocurvature perturbations which compensate for the tensor modes and 
reduce the temperature fluctuations on large scales,
which can solve the possible tension between 
the Planck observation of the temperature fluctuations~\cite{Ade:2013uln} 
and the BICEP2 detection of the B-mode polarization~\cite{Ade:2014xna},
as pointed out in~\cite{Kawasaki:2014lqa}.
We also note that
a negative tensor-to-scalar ratio $r\simeq -0.2$ may give a better fit to the Planck data~\cite{Smith:2014kka}. 
From eq.~(\ref{eq:temp_aniso2}),
we can realize such a situation when the following relation is satisfied for $\tilde{r}=-0.2$,
\begin{eqnarray}
   4 \frac{{\cal P}_{S_m}}{{\cal P}_{\zeta}}+ 
   4 \frac{{\cal P}_{\zeta S_m}}{{\cal P}_{\zeta}} + \frac{5}{6}r = \frac{5}{6}\tilde{r},
   \label{eq:negative_tensor}
\end{eqnarray}
where $\tilde{r}$ is an effective tensor-to-scalar ratio.
The left-hand side depends on $\sigma_*$, $T_{{\rm reh}}/T_{{\rm dec}}$ and $\epsilon_*$.
We plot the condition to realize the negative tensor-to-scalar ratio 
for each value of $\tilde{r}$ in Fig~\ref{fig:negative_tensor}, where we assume $\epsilon_*=0.01$.

\section{Summary and Discussions}
\label{sec:conclusion}

In this paper, we have investigated the sneturino curvaton model 
in which the CDM/baryon isocurvature perturbations are generated.
We have considered cases where the isocurvature perturbations compensate each other 
or compensate for the tensor contributions to the temperature anisotropies, 
which could improve agreement between observations.
In our curvaton scenario, 
the non-thermal leptogenesis from the sneutrino curvaton is responsible for the baryon asymmetry and 
gravitinos produced during the reheating or their decay products 
account for dark matter.
We have found that the compensation requires 
the curvaton field value during inflation of $O(10^{17})$~GeV, 
the reheating temperature $T_{\rm reh}$ of $O(10^{9\mathchar`-10})$~GeV and 
the curvaton decay temperature $T_{{\rm dec}}$ of $O(10^{6\mathchar`-7})$~GeV. 

Finally, let us comment on mass spectra of SUSY particles,
based on the constraint from the BBN~\cite{Kawasaki:2008qe,Moroi:1993mb,de Gouvea:1997tn}.
When the gravitino is the LSP, 
the gravitino mass is of $O(0.1\mathchar`-1)$~TeV.
Decay of the next-to LSP (NLSP) might destroy the success of the BBN.
If the NLSP is the left-handed sneutrino, however, 
such a problem could be avoided 
since the left-handed sneutrino dominantly decays into particles 
whose interactions are weak~\cite{Kawasaki:2008qe}. 

On the other hand,
when the LSP is not the gravitino, the gravitino may be far heavier than the LSP.
If the gravitino mass is larger than $O(10)$~TeV, the constraint from the BBN is avoided.
To obtain the observed dark matter density,
$m_{{\rm LSP}}\simeq O(0.1\mathchar`-1)$~TeV$\ll m_{3/2}$ is required. 
Such hierarchy is naturally explained 
if gaugino mass is generated only 
by the anomaly mediation~\cite{Randall:1998uk,Giudice:1998xp}, 
as is the case with 
high scale SUSY breaking models~\cite{Wells:2004di,Ibe:2006de,Hall:2011jd,Ibe:2011aa}.

\section*{Acknowledgments}
This work is supported by Grant-in-Aid for Scientific research 
from the Ministry of Education, Science, Sports, and Culture
(MEXT), Japan, No. 25400248 (M.K.), 
World Premier International Research Center Initiative
(WPI Initiative), MEXT, Japan,
and the Program for the Leading Graduate Schools, MEXT, Japan (T.H.).
K.H. acknowledges the support by JSPS Research Fellowship for Young Scientists.


\end{document}